\documentclass[aps,pra,reprint,groupedaddress]{revtex4-2}
\usepackage{graphicx}  
\usepackage{amsmath,amssymb}
\usepackage{xcolor}
\usepackage{hyperref}
\usepackage[normalem]{ulem} 

\DeclareMathOperator*{\SumInt}{%
\mathchoice%
  {\ooalign{$\displaystyle\sum$\cr\hidewidth$\displaystyle\int$\hidewidth\cr}}
  {\ooalign{\raisebox{.14\height}{\scalebox{.7}{$\textstyle\sum$}}\cr\hidewidth$\textstyle\int$\hidewidth\cr}}
  {\ooalign{\raisebox{.2\height}{\scalebox{.6}{$\scriptstyle\sum$}}\cr$\scriptstyle\int$\cr}}
  {\ooalign{\raisebox{.2\height}{\scalebox{.6}{$\scriptstyle\sum$}}\cr$\scriptstyle\int$\cr}}
}
\newcommand{\stkout}[1]{\ifmmode\text{\sout{\ensuremath{#1}}}\else\sout{#1}\fi}

\begin{document}

\title{Analytical model for attosecond time delays and Fano's propensity rules in the continuum}

\author{D.I.R. Boll}
\email[]{boll@ifir-conicet.gov.ar}
\affiliation{Instituto de Física Rosario, CONICET-UNR, Blvd. 27 de Febrero 210 bis, 2000 Rosario, Argentina}

\author{L. Martini}
\affiliation{Instituto de Física Rosario, CONICET-UNR, Blvd. 27 de Febrero 210 bis, 2000 Rosario, Argentina}

\author{O.A. Fojón}
\affiliation{Instituto de Física Rosario, CONICET-UNR, Blvd. 27 de Febrero 210 bis, 2000 Rosario, Argentina}
\affiliation{Escuela de Ciencias Exactas y Naturales, FCEIA, Universidad Nacional de Rosario, Argentina}

\date{\today}

\begin{abstract}
Extracting single photoionization time delays associated with atomic (or molecular) species from attosecond time scale two-photon experiments usually relies on the theoretical description of continuum-continuum transitions. The available models for those processes predict a universal phase contribution, independent of the angular quantum numbers of final states. However, a recent experimental-theoretical study [Fuchs, \emph{et al.} Optica 7, 154 (2020)] determined a sizable time delay dependence on the angular momentum of near-threshold photoelectrons. In this study, we present an analytical model for the two-photon two-color transition matrix amplitudes that reproduces the phase dependence on the angular quantum number of final states. Finally, we show that our analytical model can also describe the generalized Fano’s propensity rules [Busto, \emph{et al.} Phys. Rev. Lett.  123, 133201 (2019)] for laser-assisted photoionization.

\end{abstract}

\maketitle

\section{Introduction\label{sec:intro}}
The pioneering work by Eisenbud, Wigner, and Smith (EWS) on interpreting the energy derivative of the phase of the wavefunction resulted in the concept of time delays associated with quantum processes \cite{Wigner1955, Smith1960, deCarvalho2002}. For many years, that concept remained a theoretical curiosity because its experimental determination was not feasible: no measurement protocol allowed to extract time shifts on the sub-femtosecond scale. However, that barrier was removed in 2001 when attosecond science became a reality \cite{Hentschel2001, Paul2001}. Since then, techniques such as streaking \cite{Hentschel2001} or the reconstruction of attosecond beating by two-photon transitions (RABBITT) \cite{Veniard1996, Paul2001} are the methods of choice for triggering and monitoring electron dynamics in its intrinsic attosecond time scale \cite{Goulielmakis2010}. Indeed, in 2010 Schultze \emph{et al.} \cite{Schultze2010} applied the streaking protocol to determine the single photoionization (relative) time delay of electrons emerging from the $2s$ and $2p$ shells in neon atoms. Later on, Klunder \emph{et al.} \cite{Klunder2011} obtained the relative time delays for electrons ionized from the $3s$ and $3p$ shells in argon atoms, employing the RABBITT scheme.

The experimental realization of a longstanding theoretical prediction prompted a new and still active research area within the ultrafast community that ultimately led to a deeper understanding of the role played by time in quantum mechanics. In fact, as the streaking and RABBITT schemes belong to the set of pump-and-probe measurement protocols, it was soon understood that information about the EWS time delay is only accessible if the influence of the infrared probe field is considered \cite{Nagele2011}. Further theoretical developments show that observed time delays in RABBITT experiments comprise two contributions: an EWS-like delay associated with the pump stage and the additional delay originated from the probing process \cite{Dahlstrom2013}. 

These theoretical developments were crucial for the successful obtention of (relative) EWS time delay from RABBITT traces, corresponding to different atomic \cite{Klunder2011} and molecular \cite{Huppert2016} species. However, the asymptotic approximations on which they rely are equivalent to neglecting the effects of the centrifugal potential on intermediate and final states. Consequently, the continuum-continuum (cc) phase for different final states becomes a single contribution, independent of the angular quantum numbers of those states. This \emph{universal} character of the measurement-induced phases should hold as long as the pump stage generates rapidly escaping electrons. In such a scenario, the most relevant contributions to one-photon cc transitions come from space regions where short-range potentials, such as the centrifugal one, are masked by the long tail of the Coulomb interaction between the photoelectron and the parent ion. 

The above qualitative argument will be no longer valid for near-threshold photoelectron kinetic energies.  Indeed, recent measurements for slow photoelectrons \cite{Fuchs2020} confirmed the breakdown of the universal character of the delay induced during the probing stage. The theoretical analysis of these near-threshold processes \cite{Dahlstrom2013,Heuser2016,Busto2019,Fuchs2020} can be exclusively carried out through numerical simulations that, despite being extremely useful for an accurate evaluation of the observables, usually provide only partial explanations for the physical origin behind the observed effects. Therefore, the development of a complementary analytical method can be a valuable tool for providing further insights into these phenomena \cite{Armstrong2021}.

In this study, we present an analytical model that quantitatively describes the dependence of two-photon two-color transition amplitudes on the angular quantum numbers of final states. 

\section{Methodology}

\begin{figure}
 \includegraphics[width=0.75\columnwidth,clip]{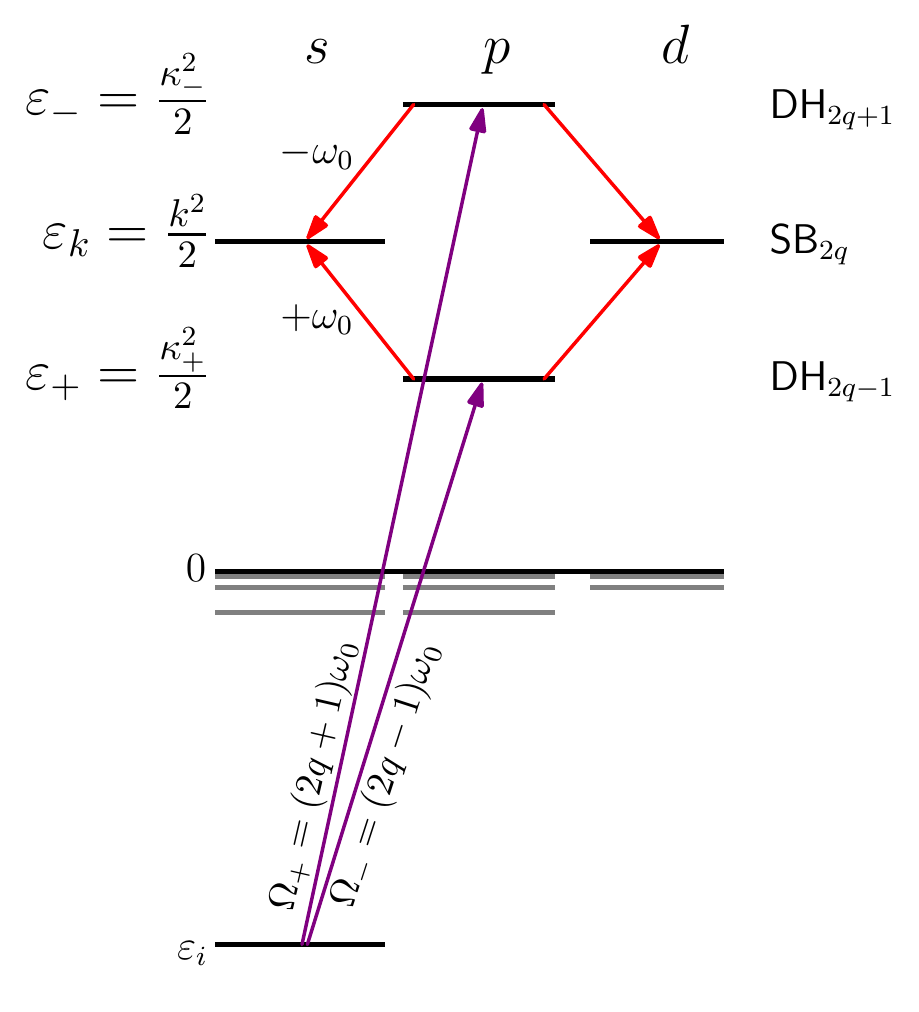}%
 \caption{(Color online) Schematic representation of pathways contributing to consecutive dressed harmonic lines and the sideband in between. For low IR intensities, dressed harmonic lines are mainly populated by transitions from ground state to continuum states, triggered by the absorption of a single photon with energy $\Omega_{\pm}=(2q\pm1)\omega_0$ from the attosecond pulse train. The interaction of these primary photoelectron distributions with the IR, gives rise to the sidebands. Time-reversed contributions, where the IR photon is absorbed or emitted first are not shown here.\label{fig:scheme}}
\end{figure}

To establish a connection with some previous experimental and theoretical results on time delays, we focus on the RABBITT scheme: atomic targets in the gas phase irradiated with a combination of odd harmonics of a Ti:sapphire laser plus a synchronized weak fraction of the fundamental infrared (IR) laser. In this situation, the photoelectron spectra display two qualitatively distinct sets of lines: a) dressed harmonic (DH) lines mainly populated by the absorption of one photon from the source of odd harmonics and b) sideband (SB) lines, populated by two-photon transitions, on which the time delay gets imprinted (see Fig. \ref{fig:scheme} for details).  

Following standard time-dependent second-order perturbation theory, it is possible to show that sideband signal reads \cite{Paul2001}, 
\begin{align}\label{eq:SB-total}
S_{2q}&\propto \sum_{L,M} \vert M_{L,M}^+ + M_{L,M}^- \vert^2,
\end{align}where $M_{L,M}^{\pm}$ are the two-photon transition matrix amplitudes connecting the initial bound state to a final state in the continuum with azimuthal and magnetic quantum numbers $L$ and $M$, respectively. In Fig. \ref{fig:scheme} we schematically show these reaction pathways for a generic sideband of order $2q$ and an initial $s$ $(l_i=0)$ state. After solving the angular algebra and the time integrals for infinitely long fields, the transition matrix amplitudes above can be factorized as
\begin{align}\label{eq:TMA}
M_{L,M}^{\pm}=A_{L,M} e^{\pm i \omega_0 \tau} \left[T_{L,\lambda}^{l_i}(\Omega_{\mp}) + T_{L,\lambda}^{l_i}(\pm\omega_0)\right],
\end{align}where $A_{L,M}$ is the angular (algebra) factor and $\exp(\pm i \omega_0 \tau)$ is a phase factor allowing to control the  sideband signal by changing the delay $\tau$ between the harmonics and the infrared laser with angular frequency $\omega_0$. The (pseudo) radial matrix elements in length gauge are given by 
\begin{align}\label{eq:RMEs}
T_{L,\lambda}^{l_i}(\Omega)=\frac{e^{i\sigma_L(k)}}{i^L}\SumInt \frac{\langle R_{\varepsilon_k,L}\vert r \vert R_{\varepsilon_{\kappa},\lambda} \rangle \langle R_{\varepsilon_{\kappa},\lambda} \vert r \vert R_{\varepsilon_i,l_i}\rangle}{\varepsilon_i+\Omega-\varepsilon_{\kappa}},
\end{align} where $\sigma_{L}(k)=\arg[\Gamma(L+1-i/k)]$ is the Coulomb phase-shift for partial wave $L$ and momentum $k$, and we include the phase factor in front of the r.h.s. of the above equation only for further convenience. The summation (integration) runs over the entire spectrum of unperturbed states of the target, with energy $\varepsilon_{\kappa}$, angular momentum $\lambda$, and radial wavefunctions $R_{\varepsilon_{\kappa},\lambda}$. The initial and final states of the system are also described by eigenstates of the field-free Hamiltonian, with radial wavefunctions $R_{\varepsilon_i,l_i}$ and $R_{\varepsilon_k,L}$, respectively.

The transition matrix amplitudes in Eq. \eqref{eq:TMA} can be derived in closed-form only for hydrogenic atoms \cite{Klarsfeld1970,Arnous1973,Jayadevan2001}. Therefore, to obtain a simple analytical model able to describe a broader range of targets, some approximations must be made. Firstly, for typical infrared laser intensities in RABBITT experiments, we can focus only on those processes where the extreme ultraviolet photon from the source of high-order harmonics is absorbed first. This is equivalent to neglect the radial matrix element $T_{L,\lambda}^{l_i}(\pm\omega_0)$ in Eq. \eqref{eq:TMA}, that accounts for the dressing of the initial state due to the infrared laser field \cite{Cionga1993}. The remaining radial matrix element can be evaluated by invoking the Dalgarno-Lewis method \cite{Dalgarno1955}, to obtain
\begin{align}\label{eq:RME-DLM}
T_{L,\lambda}^{l_i}(\Omega_{\mp})=\langle R_{\varepsilon_{k},L}\vert r\vert \rho_{\varepsilon_{\pm},\lambda}\rangle, 
\end{align} with $\rho_{\varepsilon_{\pm},\lambda}$ being the solution of the inhomogeneous differential equation $[\mathcal{H}_{\lambda}-\varepsilon_{\pm}]\rho_{\varepsilon_{\pm},\lambda}=-r R_{\varepsilon_i,l_i}$, where $\mathcal{H}_{\lambda}$ is the laser-free radial Hamiltonian for angular momentum $\lambda$, and the energy parameter satisfy $\varepsilon_{\pm}=(2q\mp 1)\omega_0+\varepsilon_i$. From the graphical representation in Fig. \ref{fig:scheme}, it can be seen that $\varepsilon_{\pm}=\kappa_{\pm}^2/2$ is merely the energy for the lower (upper) dressed harmonic line.

The theoretical developments in Ref. \cite{Dahlstrom2013} follow from approximating $\rho_{\varepsilon_{\pm},\lambda}$ and $R_{\varepsilon_k,L}$ in Eq. \eqref{eq:RME-DLM} by their asymptotic forms corresponding to large $r$. Such approximation allows to obtain analytical expressions for the radial matrix elements that factorize as $T_{L,\lambda}^{l_i}(\Omega_{\mp})=\vert T_{L,\lambda}^{l_i}(\Omega_{\mp})\vert \exp(i \phi^{\pm})$, with $\phi^{\pm}=\phi^{\pm}_{bc}+\phi^{\pm}_{cc}$. The first term $(\phi^{\pm}_{bc})$ is the phase acquired during the ionization step, triggered by the absorption of one photon with energy $\Omega_{\mp}$. The second term $(\phi^{\pm}_{cc})$ is the measurement-induced phase due to one-photon cc transitions stimulated by the infrared probe laser. One main contribution in \cite{Dahlstrom2013} is the derivation of an analytical expression for the $\phi^{\pm}_{cc}$ phases, allowing thus access to the EWS time delay. Briefly, by measuring the time offset in the oscillation of sideband signals, induced by the controlled modification of the delay $\tau$, a finite differences approximation to the EWS time delay can be obtained by subtracting the measurement-induced time delay $(\phi^{+}_{cc}-\phi^{-}_{cc})/2\omega_0$ to the total offset \footnote{Usually, in experiments, an additional phase due to the intrinsic chirp of harmonics in the train has to be compensated. One possible solution is to consider the relative time delay for electrons emerging from different atomic subshells. See Refs. \cite{Klunder2011,Huppert2016} for further details.}. 

Our derivation closely follows Ref. \cite{Dahlstrom2013} to model the radial matrix elements from Eq. \eqref{eq:RME-DLM}, but we use the exact radial wavefunction $R_{\varepsilon_k,L}$ for the final states instead of its asymptotic approximation. In that case, we get

\begin{align}
\begin{split}
T_{L,\lambda}^{l_i,\pm}\simeq&-\pi (-i)^L e^{i\sigma_L(k)}\langle R_{\varepsilon_{\pm},\lambda}\vert r\vert R_{\varepsilon_i,l_i}\rangle N_{\varepsilon_k,L} N_{\varepsilon_{\pm}}\\ &\times \int_0^{\infty} dr e^{-ikr} r^{L+2} \exp \left[i\kappa_{\pm} r+i\Phi_{\kappa_{\pm},\lambda}(r)\right]\\
& \times  \,_1F_1(L+1+i/k;2L+2;2ikr) \label{eq:our_TL_full}
\end{split}\\
\simeq &-\pi \langle R_{\varepsilon_{\pm},\lambda}\vert r\vert R_{\varepsilon_i,l_i}\rangle N_{\varepsilon_k,L} N_{\varepsilon_{\pm}} I_{k,\kappa_{\pm}}^{L,\lambda},\label{eq:our_TL}
\end{align}where the expression for the final continuum states was taken from Ref. \cite{Matsumoto1991}, $_pF_q(a_1,\ldots,a_p;b_1,\ldots,b_q;z)$ is the generalized Hypergeometric function, and the normalization factors for intermediate and final states in the continuum are given by

\begin{align}
N_{\varepsilon_{\pm}}&=\sqrt{\frac{2}{\pi \kappa_{\pm}}},\\
N_{\varepsilon_k,L}&=\frac{1}{(2L+1)!}\frac{2(2k)^L}{\sqrt{1-\exp(-2\pi/k)}}\prod_{s=1}^L \sqrt{s^2+\frac{1}{k^2}},
\end{align} respectively. The position-dependent phase in Eq. \eqref{eq:our_TL_full} is given by $\Phi_{\kappa,\lambda}(r)=\ln(2\kappa r)/\kappa+\sigma_{\lambda}(\kappa)-\pi\lambda/2$ with $\sigma_{l}(\kappa)$ the Coulomb phase-shift for partial wave $l$ and momentum $\kappa$ \cite{Dahlstrom2013}. Operating over the logarithmic phase and rearranging the factors, $I_{k,\kappa_{\pm}}^{L,\lambda}$ can be obtained in closed form by taking a limit for the integral of a closely related family of functions \cite{LandauQM}
\begin{widetext}
\begin{align}
I_{k,\kappa}^{L,\lambda}=&(2\kappa)^{i/\kappa} e^{-i\pi(\lambda+L)/2} e^{i\sigma_L(k)}e^{i\sigma_{\lambda}(\kappa)} \lim_{\epsilon \to 0^+} \int_0^{\infty} dr e^{-(\epsilon + ik-i\kappa )r} r^{L+2+i/\kappa} \,_1F_1(L+1+i/k;2L+2;2ikr)\\
\begin{split}\label{eq:integ_our_TL}
=&(2\kappa)^{i/\kappa}(-i)^L e^{-i\pi\lambda/2} e^{i\sigma_L(k)}e^{i\sigma_{\lambda}(\kappa)} \frac{\Gamma(L+3+i/\kappa)}{(ik-i\kappa)^{(L+3+i/\kappa)}} \,_2F_1\left(L+1+\frac{i}{k},L+3+\frac{i}{\kappa};2L+2;\frac{2ik}{ik-i\kappa+0^+}\right).
\end{split}
\end{align}
\end{widetext}

As the other factors entering Eq. \eqref{eq:our_TL} are strictly real \cite{Dahlstrom2013}, the bound-continuum phase acquired during the ionization step and the measurement induced phase, for two-photon processes as those depicted in Fig. \ref{fig:scheme}, must be encoded in the expression for $I_{k,\kappa}^{L,\lambda}$ above. As a matter of fact,  from Eq. \eqref{eq:integ_our_TL} we can recognize the $\phi^{\pm}_{bc}$ phase in the $ \exp{[i\sigma_{\lambda}(\kappa)]}$ phase factor and, therefore, our analytical model allows to disentangle the EWS phase contribution while the cc phases depend on the angular momentum of final states.
 
It should be noted that our model reduces to the model of Dahlstr\"om \emph{et al.} \cite{Dahlstrom2013} when the exact radial wavefunction $R_{\varepsilon_k,L}$ for the final states is replaced by its asymptotic form corresponding to large $kr$ values. Once that operation is performed, the obtention of the two-photon radial matrix elements follows along the same line as in \cite{Dahlstrom2013}.

\section{Continuum-continuum phases}
The model we present can be applied, in principle, to approximate the two-photon transition matrix amplitudes from any initial state of atomic \cite{Klunder2011} or molecular species \cite{Huppert2016,Martini2021}. This goal can be achieved by replacing the bound-continuum contributions in Eqs. (6) and (10) with the corresponding numerical results obtained from single-center expansion calculations \cite{Huppert2016}. However, to facilitate the comparison with previous theoretical and experimental data, we will firstly concentrate on initial hydrogen $s$ states. In that case, the dipole selection rules indicate that only final states with angular momentum $L=0,2$ can be populated. 

\begin{figure}
 \includegraphics[width=\columnwidth,clip]{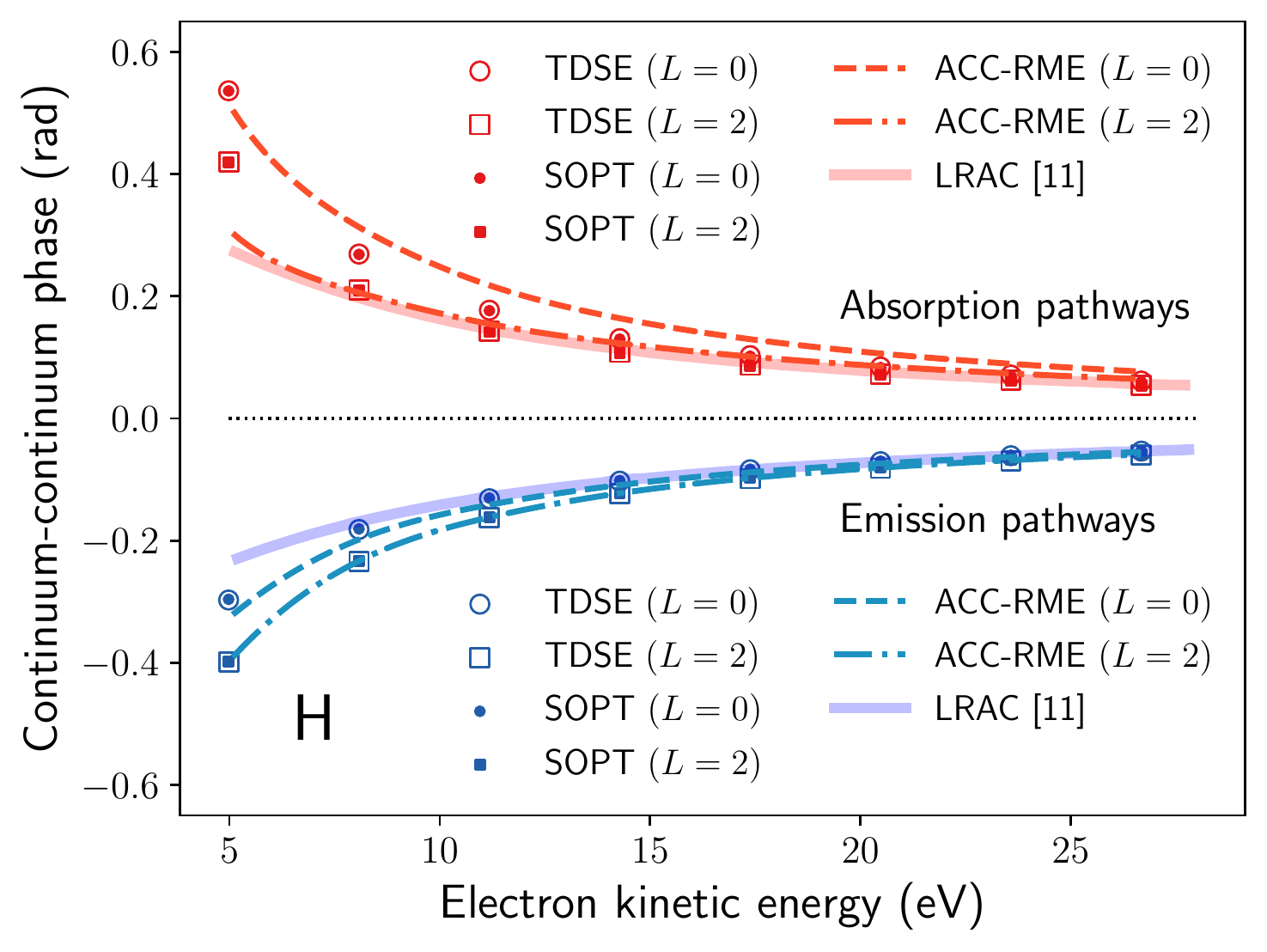}%
	\caption{(Color online) Dependence of cc phases on photoelectron kinetic energy, as obtained from different methods for hydrogen atom. Empty circles $(L=0)$ and squares $(L=2)$: TDSE. Filled circles and squares: SOPT. Long-range amplitude-corrected results are given by the full thick lines. ACC-RME results are represented by the dashed lines $(L=0)$ and the dashed-dotted lines $(L=2)$.  \label{fig:main-results}}
\end{figure}

The results we get for $\phi_{cc,L}^{\pm}$ from our analytic continuum-continuum radial matrix elements (ACC-RME) in Eq. \eqref{eq:integ_our_TL}, are shown in Fig. \ref{fig:main-results} by the dashed $(L=0)$ and dash-dotted $(L=2)$ lines. From extensive testing, we have found that our model works slightly better if momentum for intermediate states is approximated as $\kappa_{\pm}\simeq k \mp \omega_0/k$. Following this procedure, related to the soft-photon approximation \cite{Dahlstrom2013} for asymptotic intermediate states, does not change any of the conclusions reached below. To assess the accuracy of our analytical model, we additionally perform two sets of \emph{ab-initio} calculations. Firstly, we numerically solve the time-dependent Schrodinger equation (TDSE) in the velocity gauge for hydrogen atoms, employing the \textit{Qprop} code \cite{Tulsky2020}. Using a single harmonic and the IR, we obtain the transition amplitudes $M_{L,M}^{\pm}$ for the absorption and emission pathways separately. Besides, we calculate the same two-photon transition matrix amplitudes in length gauge from second-order perturbation theory (SOPT), based on the analytical results in Ref. \cite{Jayadevan2001}. For both approaches, we obtain the cc phase associated to each transition amplitude by subtracting the contributions $\phi^{\pm}_{bc}$ and $\pm\omega_0 \tau$ to the total phase \footnote{For comparison purposes, ACC-RME and TDSE results with the same kinetic energy, were shifted by a global phase.}. As can be clearly seen in Fig. \ref{fig:main-results}, the results we get for the \emph{ab-initio} calculations \footnote{The data that support the findings of this study is publicly available on Repositorio Institucional CONICET Digital \url{http://hdl.handle.net/11336/156629}} are virtually identical to each other, ruling out any gauge and/or convergence systematic errors.

Generally speaking, ACC-RME results in Fig. \ref{fig:main-results} for the emission pathways are better than those for the absorption ones. 
This behavior is not surprising and reflects a diminished descriptive power of our model when it comes to representing absorption pathways. From Fig. (1), it is evident that intermediate states for emission pathways will always have a larger kinetic energy when compared to the sideband final state. In that case, description of the intermediate states with asymptotic wavefunctions qualitatively captures the kinematical conditions under which every model is expected to work better: the lower the kinetic energy, the better the quality of wavefunctions describing that state should be. The situation is reversed for the absorption pathways. Now, intermediate states with lower kinetic energy are described through approximate asymptotic states, which may be a poor description for sufficiently low kinetic energies.

Nonetheless, ACC-RME results for the absorption pathway and $L=2$ quantitatively reproduce the accurate data for electron kinetic energies above $\varepsilon_k \sim 8$ eV, being almost identical to those from the long-range amplitude-corrected (LRAC) asymptotic approximation in Ref. \cite{Dahlstrom2013}. On the other hand, the $L=0$ results for the absorption pathway slightly overestimate \emph{ab-initio} calculations, with the notable exception of results around $E_k\sim 5$ eV, that are almost perfectly described. For the emission pathway, the quantitative agreement between the \emph{ab-initio} and ACC-RME results in Fig. \ref{fig:main-results} is prominent, considering the simplicity of our approach. For low kinetic energy, our model faithfully reproduce the phase splitting obtained through the exact calculations, whereas at high energy the results for $L=0,2$ converge to the expected limit given by the LRAC approximation. 

The above comparison of individual cc phases $\phi_{cc,L}^{\pm}$ obtained through different methods is only possible by theoretical analysis. From the experimental point of view, phase differences are the most detailed pieces of information that angular distributions of RABBITT traces encode \cite{Heuser2016,Fuchs2020}. Recently, these phase differences were experimentally obtained for helium targets \cite{Fuchs2020}, and the simulations reported therein allow to conclude that only single-active electron dynamics are triggered within this energy range, in agreement with previous calculations for two-photon two-color processes \cite{Heuser2016,Boll2020}. Therefore, we can directly compare the phase differences $\phi_{cc,0}^{\pm}-\phi_{cc,2}^{\pm}$ obtained from our model with those from accurate calculations and the available experimental data. 

\begin{figure}
 \includegraphics[width=\columnwidth,clip]{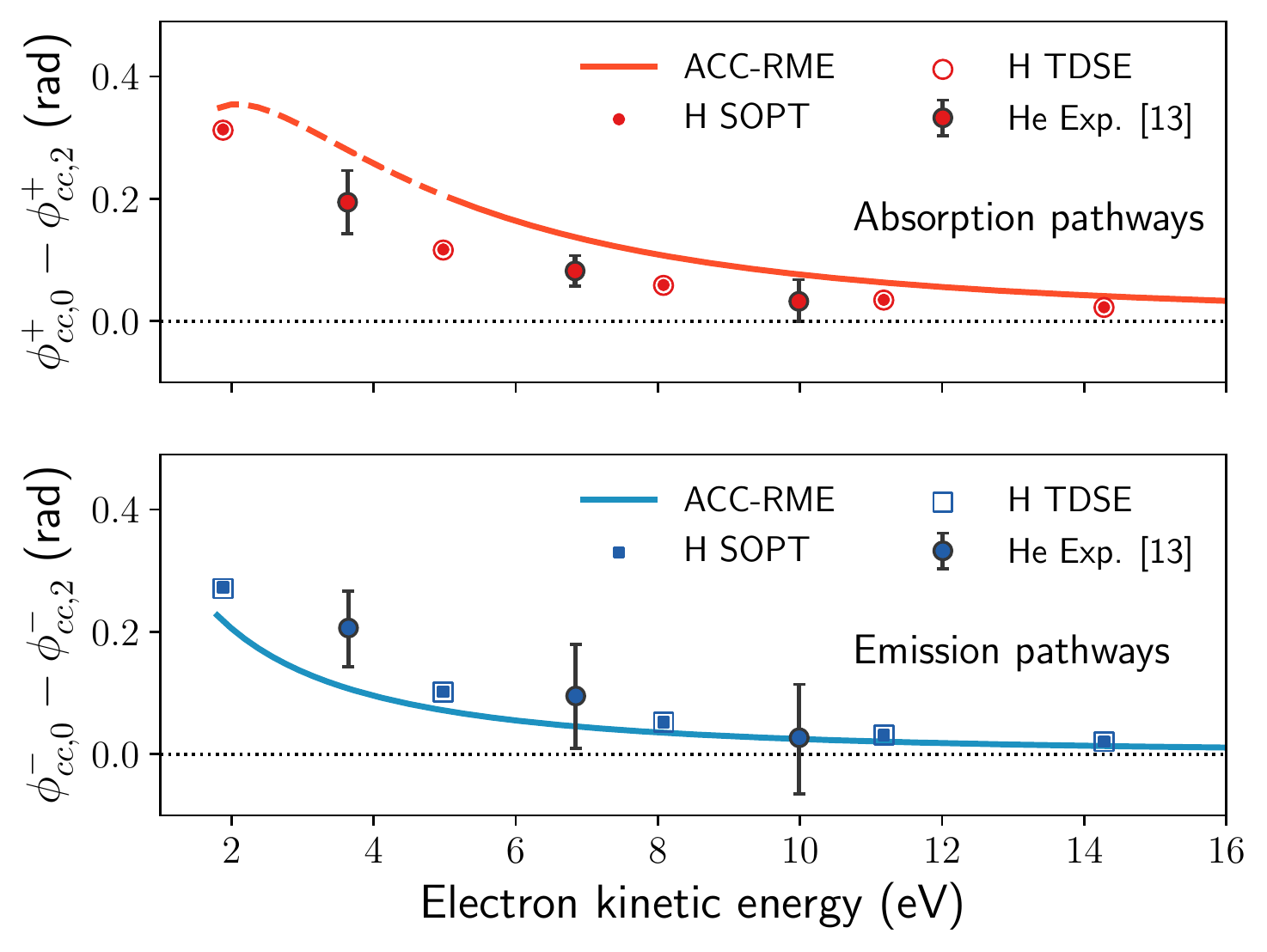}%
	\caption{(Color online) Dependence of cc phase differences on photoelectron kinetic energy. ACC-RME, SOPT and TDSE results correspond to atomic hydrogen targets. The experimental results correspond to helium atoms \cite{Fuchs2020}. Upper panel: absorption process. Lower panel: emission process. \label{fig:compara-exp}}
\end{figure}

These comparisons are presented in Fig. \ref{fig:compara-exp}, for the absorption and emission processes separately. Our model globally captures the salient features displayed by the phase differences obtained from \emph{ab-initio} and experimental data: they decrease to eventually zero for larger photoelectron kinetic energies, and, for the same kinetic energy, the results for the absorption pathways are slightly larger than those for the emission ones. The convergence to zero can be easily demonstrated by recalling that our model reduces to that in Ref. \cite{Dahlstrom2013} when final states can be accurately described by its asymptotic forms corresponding to large $kr$ values. In that case, continuum-continuum phases are independent of the angular momentum of final states and, therefore, their difference is identical to zero. Alternatively, this limit can be obtained by just replacing the Gauss hypergeometric function in Eq. \eqref{eq:integ_our_TL} with the asymptotic approximation $_2F_1(a,b;c;z) \sim \lambda_1(-z)^{-a}$, for $\vert z \vert \to \infty$ \cite{AyS1964}. For this second approach, we also obtain cc phases identical to those in Eq. (22) from Ref. \cite{Dahlstrom2013}. On the contrary, the comparatively larger values obtained for the absorption pathway cannot be easily demonstrated as this feature heavily relies on fine details of the Gauss hypergeometric function. 

As before, we observe a better agreement for the emission pathway where the ACC-RME results quantitatively reproduces the experimental data and \emph{ab-initio} simulations, for electron kinetic energies above $\sim 5$ eV. For smaller kinetic energies, our model captures the trend set by the simulations for hydrogen and the experimental data for helium. The results for the absorption pathway are less accurate, as ACC-RME calculations overestimate both the experimental data and the \emph{ab-initio} calculations on the entire range of electron kinetic energies under analysis. Again, the model can reproduce the trend showing an increase in the phase difference for smaller electron kinetic energies. The results for electron kinetic energies below $\sim 5$ eV are shown in dashed line to emphasize that for an 800 nm infrared laser, the intermediate states of the absorption pathway will have a kinetic energy below $\sim 3.5$ eV, a value perhaps too low to admit a description with asymptotic states.

\section{Analysis of generalized Fano’s propensity rules in the continuum}

The primary aim of developing this model was to put forward a general analytical expression able to reproduce phase differences observed for partial waves populating the final channel in sidebands. The previous analysis demonstrates that this goal is achieved. In addition, as a by-product, with our model we can also study the branching ratio for partial waves upon cc transitions. This concept can be traced back, at least, to a few experimental and theoretical studies for the half-RABBITT sidebands obtained when atomic or molecular targets are ionized with a combination of a free-electron laser pulse and an infrared probing field \cite{Meyer2010,Fushitani2012}. More recently \cite{Busto2019}, this concept was applied to RABBITT experiments, generalized for different partial waves, and rationalized as a superset of Fano's propensity rules \cite{Fano1985}. The generalized Fano propensity rule states that upon absorption (emission) of one IR photon for the continuum-continuum transition, the partial wave with larger (smaller) angular momentum is favored \cite{Busto2019}.    

\begin{figure}
 \includegraphics[width=\columnwidth,clip]{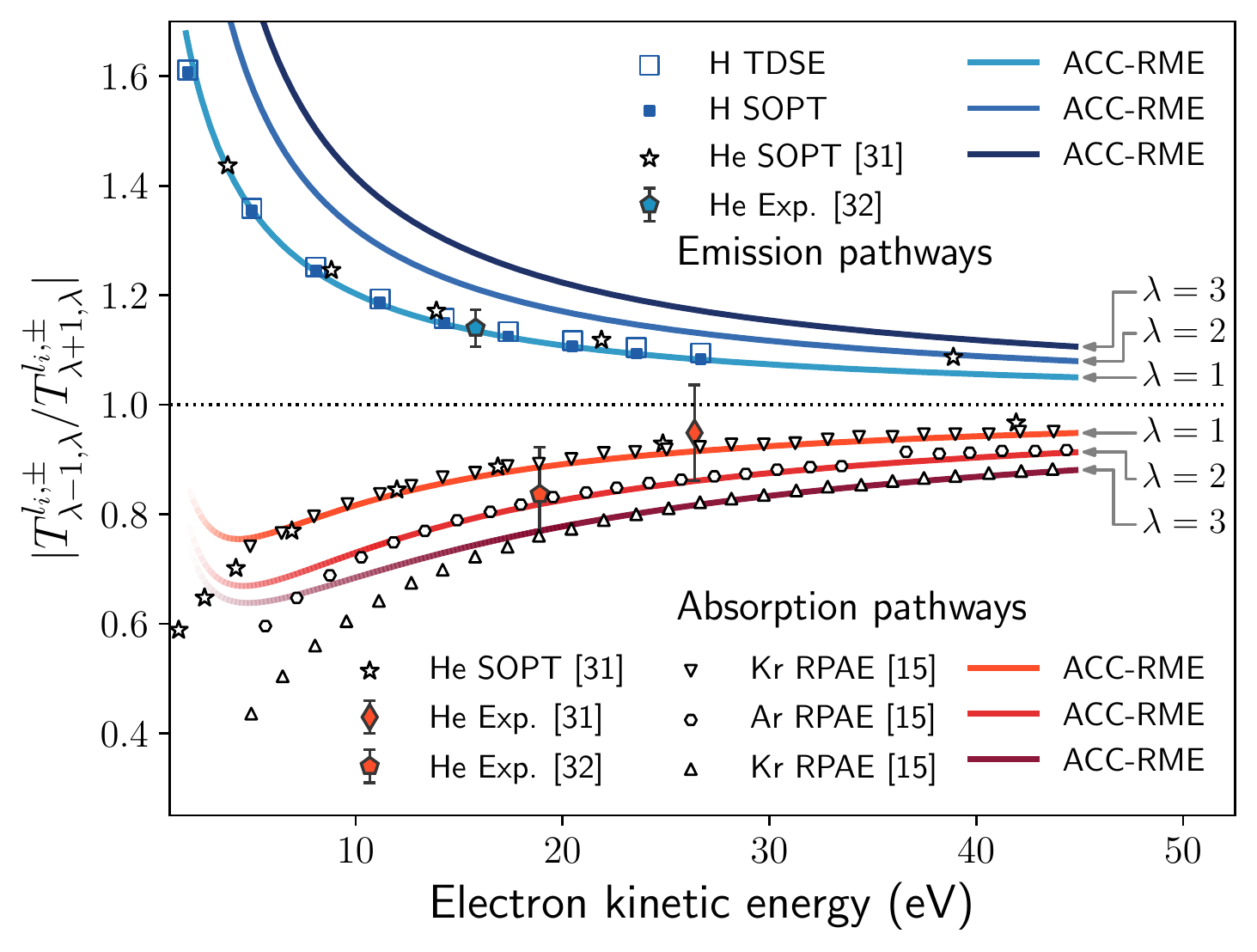}%
	\caption{(Color online) Moduli ratios for the absorption (red) and emission (blue) pathways. ACC-RME results for different intermediate states are given by full lines. Second-order perturbation theory calculations are given by empty stars \cite{Meyer2010} and full squares (emission only). TDSE simulation results are shown by empty squares (emission only). RPAE results \cite{Busto2019} for different intermediate states are shown by empty triangles and hexagons. \label{fig:moduli-compara}}
\end{figure}

In Fig. \ref{fig:moduli-compara}, we compare the moduli ratios $\vert T_{\lambda-1,\lambda}^{l_i,\pm}/T_{\lambda+1,\lambda}^{l_i,\pm} \vert $ obtained from our model with the benchmark data: random-phase approximation with exchange (RPAE) \cite{Busto2019}, TDSE and SOPT calculations and the scarce experimental data available in the literature \cite{Meyer2010,Fushitani2012}. In any case, we observe that results obtained from our model reproduce the behavior expected from the generalized Fano propensity rules. For absorption pathways, the radial matrix elements for partial waves with $L=\lambda+1$ are larger than those for $L=\lambda-1$ and, therefore, their ratio is smaller than its asymptotic value of one. For the emission pathways, the opposite trend is observed and partial waves with smaller angular momentum dominates. Even more, our analytical results show that this behavior is dictated exclusively by continuum-continuum transitions taking place in a effective potential $(Z_{net}-1)/r$, with $Z_{net}$ being the net charge of the target, irrespective of the more subtle details that its electronic structure may have.

For $\lambda=1$, the model almost perfectly describes \emph{ab-initio} and experimental results for the emission pathway on the entire energy range. Conversely, the results for the absorption pathway fail for electron kinetic energies below $\sim 5$ eV. As before, this behavior may be ascribed to a poor description of low kinetic energy intermediate states with asymptotic wavefunctions. The approximately linear dependence on $\lambda$ of the kinetic energy value below which model results deviate from the accurate calculations emphasizes the previous conclusion. Further understanding of this trend would require a direct comparison of the approximate intermediate states with those obtained from the numerical resolution of the Dalgarno-Lewis method, which is beyond the scope of the present study.    

The final remark on these branching ratios relates to its universal character. From Eq. \eqref{eq:our_TL}, it is clear that two-photon radial matrix elements factorize as the product of a radial matrix element for the pump stage and a single-photon cc factor. Therefore, the agreement of our model results with those from more accurate calculations and the experimental data for the branching ratios in Fig. \ref{fig:moduli-compara} is a direct demonstration that the observed variations are dictated by the single-photon cc transitions. In turn, it means that correlation effects, included in the RPAE calculations \cite{Busto2019} and experimental data \cite{Meyer2010,Fushitani2012}, play little to no role in these branching ratios.

The above analysis of branching ratios for the radial matrix elements can be systematically performed from a theoretical perspective. On the contrary, the empirical verification of these results has some nuances depending on the initial state. On one hand, for atomic $s$ states there will be a single intermediate state with $\lambda=1$, therefore, the experimental and theoretical results can be directly compared. The agreement we observe in Fig. \ref{fig:moduli-compara} indicates that the branching ratios are independent of the radial distribution for the initial $s$ state. On the other hand, measurements for initial states with angular momentum larger than zero will be hindered by the fact that final states with $L=l_i$ are coherently populated by contributions following two different quantum pathways. In such a scenario, a branching ratio dependence on the radial distribution of the initial state cannot be ruled out. This is in agreement with the findings in Ref. \cite{Martini2021} (Fig. 5), where changing the radial distribution for a $1s$ initial state in hydrogen does not modify an observable closely related to radial matrix elements, whereas the same procedure for initial $2p$ orbitals induces a dramatic change.    

\section{Conclusions and outlook}
To summarize, we develop an analytical model for the two-photon two-color transition matrix amplitudes that quantitatively reproduce state-of-the-art calculations and experimental data at low photoelectron kinetic energies, without resorting to fitting free parameters. We demonstrate that the effects induced by the centrifugal barrier onto the wavefunctions can be almost completely recovered by just using accurate final states, even if intermediate states are approximated by its asymptotic form corresponding to large $r$. Besides, the functional form we obtain for the matrix elements is fully compatible with the procedure to extract the EWS time delays from the sideband signal oscillation in RABBITT experiments. Additionally, our results can be readily adapted to upgrade models describing higher order cc processes \cite{Bharti2021}, as it only requires to change the description of final states. Finally, we demonstrate that the partial waves branching ratio above a $\lambda$-dependent photoelectron kinetic energy can also be accurately retrieved by considering the centrifugal potential effects only on final states. In particular, our model may be useful to disentangle the coherent contributions populating final state partial waves with $L=l_i$ in two-color experiments from free-electron laser facilities, where measurements can be performed away from threshold.

\begin{acknowledgments}
The authors acknowledge the financial support from the Agencia Nacional de Promoción Científica y Tecnológica, Project PICT No. 01912 (2015-3392), and from the Consejo Nacional de Investigaciones Científicas y Técnicas de la República Argentina, Project PIP No. 0784 (11220090101026). Part of the results presented in this work have been obtained by using the facilities of the CCT-Rosario Computational Center, member of the  High Performance Computing National System (SNCAD, MincyT-Argentina).
\end{acknowledgments}

\bibliography{cc-phases}

\end{document}